\documentclass[
reprint,
superscriptaddress,
showkeys,
amsmath,amssymb,
aps,
pra,
floatfix,
]{revtex4-1}
\usepackage{CJK}
\usepackage{subcaption}
\usepackage{bm}% bold math
\usepackage{turnstile}

\usepackage{graphicx,float}
\usepackage{amsmath,amssymb}
\usepackage{url}
\usepackage{caption}

\usepackage{epsfig}

\begin{document}
%S\listoffigures
\captionsetup[figure]{labelformat={default},labelsep=period,name={FIG.}}

\begin{CJK*}{GB}{} % Use default fonts from CJK (see below)

\title{
Interaction of a wave with an accelerating object and the equivalence principle
}
\author{A.~I.~Frank}
\email[frank@nf.jinr.ru] \\
\affiliation{Joint Institute for Nuclear Research, 141980 Dubna, Russia}

\date{\today}

\begin{abstract}
A new look at the so-called effect of an accelerating matter is presented. It was previously stated that the effect is optical in nature and consists in changing the frequency of the wave passing through a refractive sample moving with acceleration. However, from a simple consideration based on the principle of equivalence, it follows that the idea of the connection of the effect only with the refraction phenomenon is unreasonably narrow, and a change in the wave frequency should inevitably occur during scattering by any object moving with acceleration. Such an object can be either an elementary scatterer or any device transmitting a narrowband signal.
\end{abstract}


\maketitle
\end{CJK*}

%\section{}
For the first time the phenomenon that later became known as the accelerating matter effect was briefly mentioned in the PhD thesis by V.I. Mikerov~\cite{Mikerov77}. Analyzing the possibility of filling a trap for ultracold neutrons (UCN) without its depressurization, he proposed using a thin membrane moving in harmonic law with and against the UCN motion. At the same time, he found that the energy of UCNs, after their passing through an oscillating foil, should change. Being unpublished, this result has remained unknown for a long time.

In 1982 K. Tanaka~\cite{Tanaka82} found a solution to the problem of an electromagnetic wave passing through a linearly accelerating dielectric plate and predicted that the frequency of a wave transmitted through such a sample differs from the frequency of an incident wave. With a single passage of a wave through the plate, the frequency change is determined as follows:
\begin{equation}\label{eq1}
\Delta\omega\cong\frac{\omega ad}{c^{2}}(n-1), \quad \left(\frac{ad}{c^{2}}\ll1\right),
\end{equation}
where $\omega$ is the incident wave frequency, $n$ is the refraction index, $d$ is the plate thickness, $a$ is the acceleration, and $c$ is the speed of light.  The possibility of observing this optical effect was discussed in~\cite{Neutze98}, but, as far as we know, due to the smallness of the effect, the experiment was never performed.
In 1993 the work of F.V. Kowalski~\cite{Kowalski93} appeared, where he proposed to verify the equivalence principle in a new type of neutron experiment. His theoretical approach was based on the concept of the group and phase velocities of the neutron. He also calculated the time of wave propagation between the corresponding points in the laboratory coordinate system and in a system moving with acceleration. In both cases, the wave passed through a refracting sample. As an intermediate result, the author concluded that the neutron energy changes as it passes through a sample moving with a not too much acceleration. The energy change was determined as:
\begin{equation}\label{eq2}
\Delta E=mad\left(\frac{1-n}{n}\right),
\end{equation}
where $m$ is the neutron mass. 

Later, the same issue was considered by V.G. Nosov and A.I. Frank~\cite{Nosov98}. The analysis aimed at sequential calculation of the velocity of neutrons at their entering the sample, propagating in the medium, and escaping through another surface, was in fact based on the classical approach. To estimate the magnitude of the effect, the authors obtained a formula that coincided with the result by F.V. Kowalski~\eqref{eq2}.

It should be noted that in~\cite{Nosov98} the formula~\eqref{eq2} was obtained under two important assumptions. Firstly, it was assumed that the quantum effects are small, whereas according to the second assumption, the motion of the substance where the neutron wave propagates does not affect the result, and all changes in the wave function are caused only by the motion of the  sample boundaries.

The first short report on the experimental observation of changes in the energy of neutrons as they pass through an accelerating sample appeared in 2006~\cite{Frank2006}. Soon, the experiment was repeated under better conditions~\cite{Frank2008}. In both cases, the change in the energy of UCNs passing through a sample oscillating in space was measured. The time-varying acceleration of the sample reached values of $60-70$ m/s$^2$. For an energy transfer of  $(2-6)\times 10^{-10}$~eV, a good agreement with the formula~\eqref{eq2} was obtained.

Afterwards, acceleration and deceleration of neutrons passing through an oscillating plate of a refractive material were observed in an experiment sensitive to neutron velocity~\cite{Frank2011}. The obtained results turned out to be in satisfactory agreement with the expected ones.

A certain development was given to the subject of physical interpretation of the effect. The optical approach to the description of its nature dating back to~\cite{Tanaka82} naturally appealed to the Doppler shift in the wave frequency upon refraction into a moving medium~\cite{Frank2013}. In this case, the frequency of a neutron wave in a moving medium measured in the laboratory system is:
\begin{equation}\label{eq3}
\omega_{i}=\omega_{0}+(n'-1)k_{0}V, \quad (V\ll v_{0}),
\end{equation}
where $\omega_{0}$ and $\omega_{i}$ are the frequencies of the incident and the refracted wave, respectively, $n'$ is the wave refraction index in the coordinate system of a moving substance, $v_{0}$ and $k_{0}$ are the velocity and the wave number of the incident wave, and V is the velocity of the medium and its boundary. 

If a neutron wave passes through a uniformly moving layer of matter, the Doppler frequency shifts that occur when a wave goes through its two boundaries are equal in magnitude but opposite in sign. The total effect of frequency change is equal to zero. In the case of accelerated motion, the frequency shifts on the entrance and escape surfaces are different, since during the wave propagation through the sample, the velocity of the boundary changes by $\Delta V=a\tau$, where $a$ is the acceleration, and $\tau$ is the sample transit time. In the case of a refracting sample, the difference between the frequency of the incident and transmitted neutron waves is:
\begin{equation}\label{eq4}
\Delta\omega=ad \left(\frac{1-n}{n}\right) \frac{k_{0}}{v_{0}}=\frac{mad}{\hbar} \left(\frac{1-n}{n}\right), \quad (V, at\ll v_{0}),
\end{equation}
and the energy change $\Delta E=\hbar\Delta\omega$ is determined by~\eqref{eq2}. 
 Recently, the authors of\cite{Voronin2014,Braginetz2017} observed a change in the energy of cold neutrons passing through an accelerating sample under conditions close to those of the Bragg reflection. Explaining the nature of the effect, they resorted to the facts very similar to those given above. It seems obvious that the energy change effect observed in these works is, in essence, of the same nature as the accelerating matter effect at refraction, although the details of the phenomenon are mostly different.
 
The optical approach to describing the effect proved to be productive, and, therefore, it was subsequently extended onto the case of a birefringent material. In the case of neutron optics, the phenomenon of birefringence, well-known in conventional optics, is based on the spin dependence of the refractive index. In~\cite{Frank2013}, it was shown that if a neutron goes through a birefringent substance moving with acceleration, it results in occurrence of a non-stationary state with a precessing spin. A similar situation can happen when a two-component neutrino passes through a layer of accelerating matter. In this case, the wave function of the resulting state changes in such a way that significantly alters the nature of the neutrino state evolution at the subsequent propagation in free space.

However, the existence of the accelerating matter effect (AME) can also be assumed without relying on specific optical calculations but based only on the equivalence principle (EP). To demonstrate this, the authors of~\cite{Frank2008} turned to Kowalski's Gedanken experiment~\cite{Kowalski93}, but unlike him, they proceeded from the idea of an undoubted validity of the EP. Hereafter, their argumentation will be presented, as it is crucial for what comes next.

\begin{figure*}[htb] 
\centering
%\includegraphics[width=\columnwidth]{Fig1.eps}
%\includegraphics[width=2\columnwidth]{Fig1.eps}
%\captionsetup[subfigure]{labelformat=simple}
    %\hspace{-3\baselineskip}
    \begin{subfigure}[b]{0.5\columnwidth}
        \includegraphics[width=\textwidth]{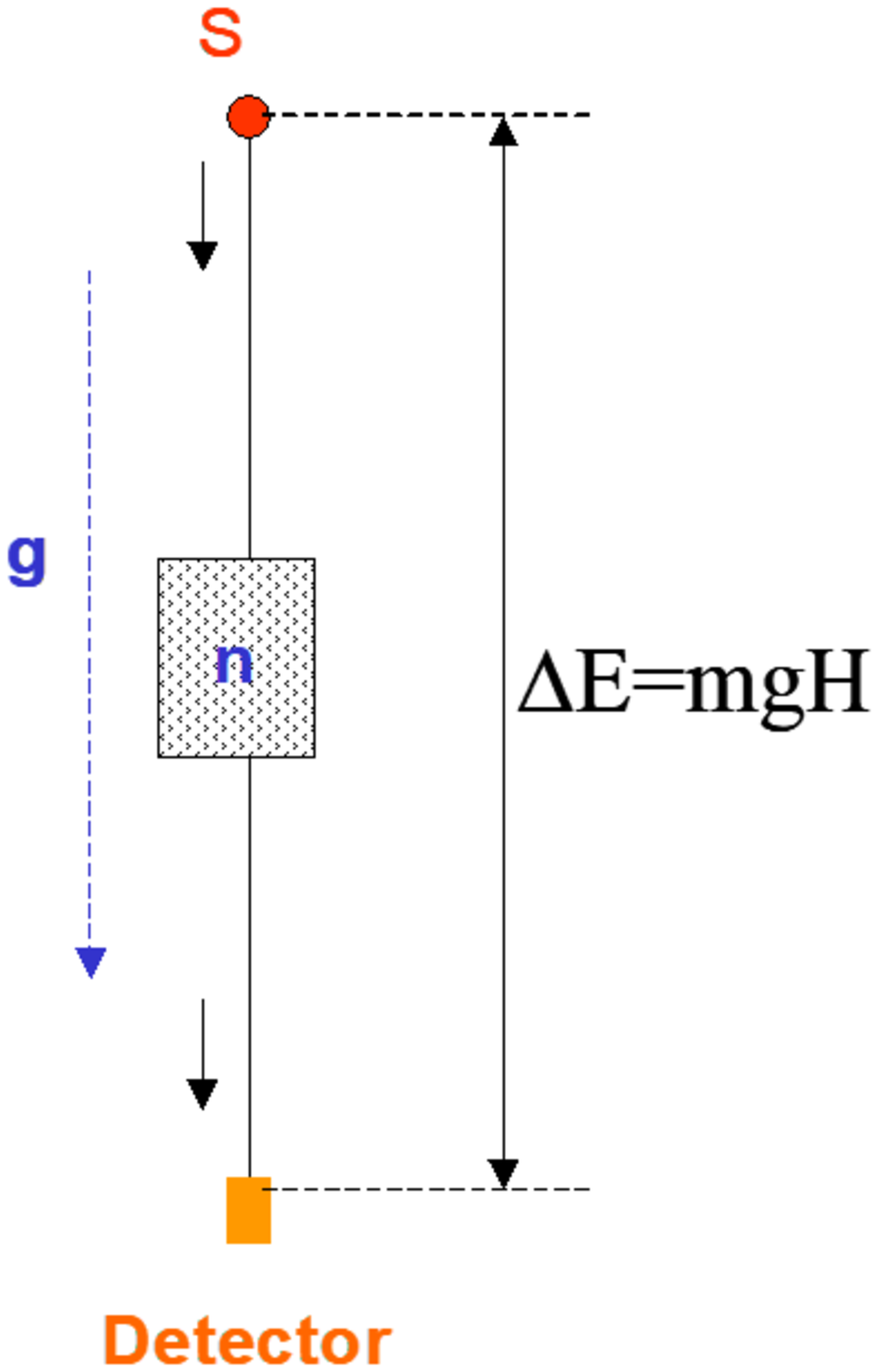}
        \label{fig:fig1a}
        \vspace{-1.5\baselineskip}
        \caption{}
    \end{subfigure}
   \hspace{3\baselineskip}
    \begin{subfigure}[b]{0.7\columnwidth}
        \includegraphics[width=\textwidth]{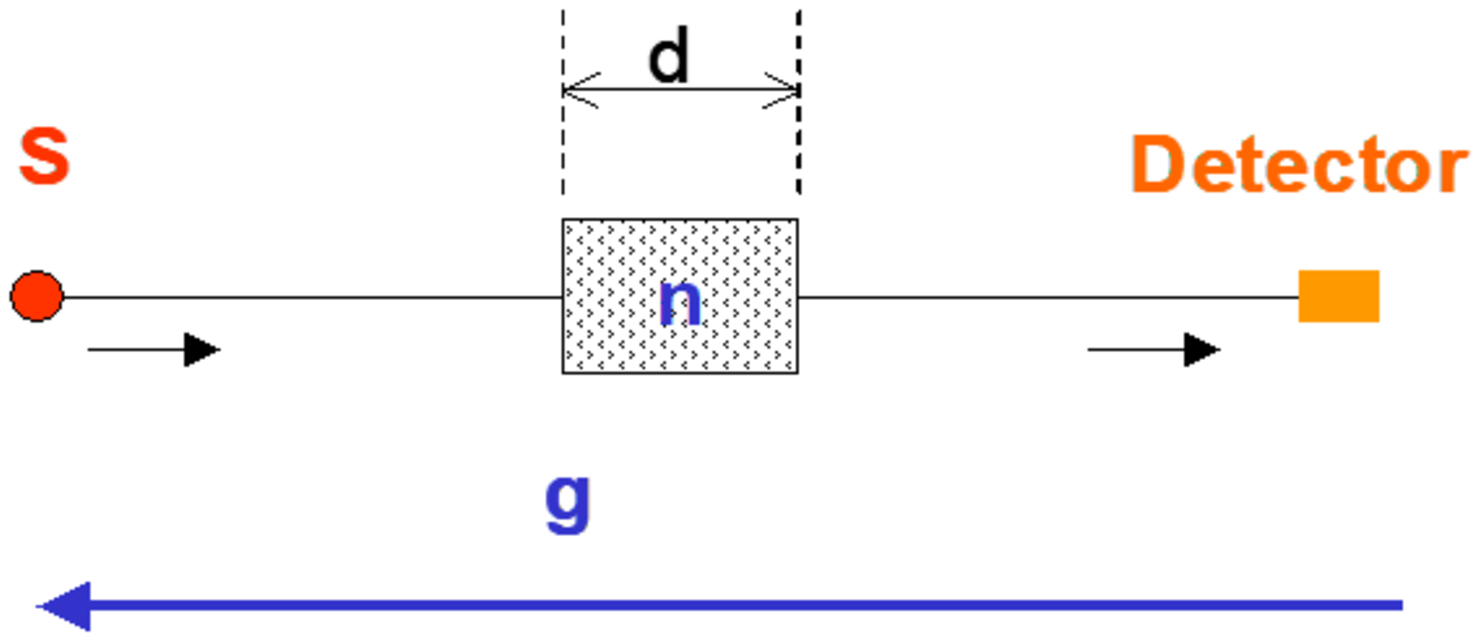}
        \label{fig:fig1b}
         \vspace{-1.5\baselineskip}
         \caption{}
    \end{subfigure}
\caption
{
Neutron passing through a refractive sample in the inertial system with gravity (on the left) and in the non-inertial system of coordinates
}
\label{fig1}
\end{figure*}

Figure~\ref{fig1} shows the case when a neutron moves from the source to the detector, accelerating due to the gravity. Obviously, the neutron energy at the point of detection exceeds the initial energy by $mgH$, where $H$ is the difference between the height of the source and that of the detector, and $g$ is the free fall acceleration. The introduction of a refractive slab on the neutron path does not change this conclusion. Let us now turn to the case shown on the right side of Fig.~\ref{fig1}. Here, the source, the refracting plate, and the detector move together with the same acceleration $a = g$, occurring in a non-inertial reference frame, while the neutron observed from the laboratory frame moves at a constant velocity. 

By virtue of the equivalence principle, the results of all measurements in a non-inertial reference frame must be identical to those in an inertial frame in the presence of gravity, and this applies fully to the energy measured by the detector. In the trivial case of no sample, it is unquestionable. However, if a refractive slab moving at the same acceleration as the detector is introduced on the neutron path, the neutron time of flight is delayed, since the neutron speed is less in matter than in a vacuum (to be definite, we assume $n<1$). Neglecting a small difference in $\Delta t$ for a stationary and an accelerating sample, we obtain the following:
\begin{equation}\label{eq5}
\Delta t=\frac{d}{v_{0}} \left(\frac{1-n}{n}\right),
\end{equation}
where $d$ is the sample thickness. During the delay time $\Delta t$ the detector will be accelerating. An additional change in its velocity, (in relation to the case where there is no slab) is $\Delta v=a\Delta t$.
 
Thus, if the role of a refractive slab reduced to a time delay, then the velocity of the neutron with respect to the detector at the instant of reaching the latter and, hence, the measured energy would differ from their counterparts in the absence of matter. This would contradict the equivalence principle. Therefore, the passage through an accelerated slab must be accompanied not only by a delay in time but also by a change in the neutron energy such that it compensates for the effect of the additional acceleration of the detector within the time $\Delta t$. Calculating this value in accordance with $\Delta E=mv \cdot a\Delta t$, we obtain formula~\eqref{eq2}.

Similar reasoning can be made for the case of a photon. The time delay occurring due to refraction of light in a medium
\begin{equation}\label{eq6}
\Delta t=d(n-1)/c,
\end{equation}
where $c$ is the speed of light. The Doppler shift in the frequency caused by the accelerating detector is, in the first order by $v/c$
\begin{equation}\label{eq7}
\Delta\omega\approx\omega a \Delta t/c=\omega\Delta t/c.
\end{equation}
Applying~\eqref{eq6} to~\eqref{eq7}, we obtain Tanaka's formula~\eqref{eq1}.  

Therefore, proceeding only from the equivalence principle and neglecting both relativistic and non-stationary quantum effects, we can obtain expressions for the AME, which coincide with the results of a more rigorous theoretical analysis both for light and a slow particle. The formula~\eqref{eq7} written as
\begin{equation}\label{eq8}
\Delta\omega=ka\Delta t,
\end{equation}
where $k$ is the wave number, is valid for both cases.

It should be emphasized that in the above-mentioned considerations, we came to the conclusion about the change in the frequency of a wave passing through an accelerating sample, relying only on the assumption of different propagation times of wave propagation in the sample and in vacuum,  but in no way on the assumption that this difference is connected precisely with refraction in matter. Consequently, this conclusion will remain valid if, instead of a refracting sample, there will be an object moving with acceleration and transmitting a signal with a delay.

The variety of such objects is extremely large. It is known, for example, that in quantum mechanics interaction is inevitably associated with time delay tentatively described by the so-called group delay time (GDT),~\cite{Eisenbud48,Bohm51,Wigner55}
\begin{equation}\label{eq9}
\tau=\hbar \frac{d \phi}{dE},
\end{equation}
where $\phi$ is the interaction amplitude phase (for example, scattering) and $E$ is the energy. Thus, any elementary scatterer moving with acceleration must change the wave frequency. This is an effect that complements the usual Doppler shift and is proportional to acceleration rather than velocity.

Any transmission of a signal occurs with a time delay, therefore, an accelerated device that receives and transmits a wireless signal will inevitably transmit it at a different frequency. Such a device may be, for example, a transceiver of a radio or acoustic signal or a fiber-optic line being an alternative to a refractive plate considered by K. Tanaka. However, it is important that the element of the device receiving radiation, i.e. the receiver, would move at the same speed, as the emitter. In this case, a receiving and emitting antenna or just an end of a waveguide or a fiber line can serve as the receiver and emitter.
%---------------------------------------------------
\begin{acknowledgments}
The author expresses his sincere gratitude to G.M. Arzumanyan, V.A. Bushuev, M.A. Zakharov, G.V. Kulin, S.V. Goryunov for useful discussions.
\end{acknowledgments}

% Create the reference section using BibTeX:
\bibliography{Frank_arxiv.bib}
\end{document}